\documentclass[aps,pra,amsmath,amssymb,superscriptaddress, nofootinbib]{revtex4-1}
\usepackage{graphicx}
\usepackage{dcolumn}
\usepackage{bm}
\usepackage{bbm}
\usepackage{amsmath}

\usepackage{epstopdf}
\usepackage{graphicx}
\usepackage{dcolumn}
\usepackage{bm}
\usepackage{bbm}
\usepackage{amsmath}
\usepackage{color}
\usepackage{multirow}
\usepackage{hyperref}
\usepackage{subfigure}

\newcommand{\ket}[1]{\left|{#1}\right\rangle}
\newcommand{\bra}[1]{\left\langle{#1}\right|}
\newcommand{\braket}[2]{\langle{#1}|{#2}\rangle}

\newcommand{\aref}[1]{\hyperref[#1]{Appendix~\ref{#1}}}

\begin{document}

\title{Centrality measure based on continuous-time quantum walks and experimental realization}

\author{Josh A. Izaac}
\altaffiliation{JAI and XZ, as the first named authors, have contributed equally to this work.}
\affiliation{School of Physics, The University of Western Australia, Perth, WA 6009, Australia}

\author{Xiang Zhan}
\altaffiliation{JAI and XZ, as the first named authors, have contributed equally to this work.}
\affiliation{Department of Physics, Southeast University, Nanjing 211189, China}

\author{Zhihao Bian}
\affiliation{Department of Physics, Southeast University, Nanjing 211189, China}
\author{Kunkun Wang}
\affiliation{Department of Physics, Southeast University, Nanjing 211189, China}
\author{Jian Li}
\affiliation{Department of Physics, Southeast University, Nanjing 211189, China}
\author{Jingbo B. Wang}\email{jingbo.wang@uwa.edu.au}
\affiliation{School of Physics, The University of Western Australia, Perth, WA 6009, Australia}
\author{Peng Xue}\email{gnep.eux@gmail.com}
\affiliation{Department of Physics, Southeast University, Nanjing 211189, China}

\date{\today}

\begin{abstract}
Network centrality has important implications well beyond its role in physical and information transport analysis; as such, various quantum walk-based algorithms have been proposed for measuring network vertex centrality. In this work, we propose a continuous-time quantum walk algorithm for determining vertex centrality, and show that it generalizes to arbitrary graphs via a statistical analysis of randomly generated scale-free and Erd\H{o}s-R\'enyi networks. As a proof of concept, the algorithm is detailed on a 4-vertex star graph and physically implemented via linear optics, using spatial and polarization degrees of freedoms of single photons. This paper reports the first successful physical demonstration of a quantum centrality algorithm.

\end{abstract}

\maketitle

\section{Introduction}
Since the seminal paper by \citet{aharonov1993}, quantum walks have become a fundamental tool in quantum information theory \cite{kempe2003}, allowing us to bridge the often more esoteric world of quantum computation and algorithms \cite{childs2004, lovett2010, venegas-andraca2012, wang2014, Li2015} with real-life graph and network theory \cite{martinez-martinez2016, berry2011, douglas2008} and dynamical quantum modelling applications \cite{sarovar2010,rebentrost2009,engel2007,izaac2013}. This is due, in part, to the markedly differing behaviour of the quantum walk compared to its classical analogue. Harnessing the effects of superposition, quantum coherence and entanglement, the quantum walk propagates quadratically faster, providing a key source for new quantum algorithms and a platform for universal quantum computation \cite{childs2009, childs2013, qiang2016}.

Like the classical case, quantum walks are divided by two distinct approaches --- the discrete-time quantum walk (DTQW), which introduces spin-states and a quantum coin operation with discrete time-evolution operators, and the continuous-time quantum walk (CTQW), which evolves the walker continuously in time \cite{farhi1998}. Due to the enlarged Hilbert space and higher degrees of freedom of the DTQW, the relationship between these two formulations is inherently non-trivial; regardless, an equivalency has been explored using both a limiting approach \cite{strauch2006} and percolation theory \cite{philipp2016}.

One potential application of the quantum walk is in providing an efficient quantum algorithm for vertex centrality ranking in network analysis. Previous studies have proposed algorithms built on the standard discrete-time quantum walk~\cite{Berry2010}, the Szegedy discrete-time quantum walk \cite{paparo2012,paparo2013,loke2017}, or the continuous-time quantum stochastic walk (QSW) \cite{loke2017,sinayskiy2013,falloon2016}. However, whilst comparing well to classical centrality measures, these have the distinct disadvantage of requiring expanded Hilbert spaces (up to $N^2$ dimensions for a graph of $N$ vertices), or in the case of the QSW, muting the quantum behaviour due to decoherence.

In this study, we propose an alternative quantum walk centrality algorithm based on the CTQW, allowing us to preserve the full quantum behaviour of the walker, whilst limiting the dimension of the Hilbert space to $N$. Furthermore, we have experimentally implemented this algorithm in the case of the 4-vertex star graph. As far as we are aware, this is the first quantum centrality measure to be physically implemented.

This paper is structured as follows. In \autoref{sec:ctqw}, we describe the continuous-time quantum walk and its relationship to classical random walks. We then briefly discuss graph centrality measures in \autoref{sec:centrality}, before introducing our CTQW-based quantum centrality scheme in \autoref{sec:ctqwcentrality}. A thorough statistical analysis using ensembles of randomly generated graphs is presented in \autoref{sec:stats}, highlighting the suitability of the quantum centrality scheme for general graphs.  Then, we discuss our experimental implementation via linear optics in \autoref{sec:experiment}, before finally presenting our conclusions in \autoref{sec:conc}.

\section{Classical and quantum walks}
\label{sec:ctqw}

\subsection{Classical random walks}
Consider an arbitrary undirected graph $G(V,E)$, composed of vertices $j\in V$ and edges $(i,j)\in E$ with $|V|=N$. The adjacency matrix of $G$ is a symmetric matrix defined by
\begin{align}
	A_{ij}=\begin{cases}
		1,&(i,j)\in E\\
		0,&(i,j)\notin E
	\end{cases}
\end{align}
A discrete-time random walk (DTRW) over $G$ is a stochastic Markovian process that evolves as follows,
\begin{align}
	\mathbf{P}^{(n+1)}=T\mathbf{P}^{(n)}
\end{align}
where $\mathbf{P}^{(n)}_i$ represents the probability of finding the walker at vertex $i$ at time-step $n$, and $T$ the transition matrix. As per convention, the transition matrix is normally taken to be 
\begin{align}
	T=AD^{-1}
\end{align}
with $D_{ij}=\delta_{ij}\sum_{k}A_{ki}$ a diagonal matrix containing the vertex degrees of the graph. This ensures that $T$ is \textit{stochastic} ($\sum_k T_{ki}=1$), preserving the probability of the walker. The steady-state limiting probability distribution of the walker,
\begin{align}
\lim_{n\rightarrow\infty}T^n\bm{P}^{(0)}=\bm{\pi}
\end{align}
must satisfy the equation $T\bm{\pi}=\bm{\pi}$. Thus, the limiting-distribution is simply the eigenvector of $T$ with eigenvalue $\lambda=1$. It is trivial to show that the limiting distribution is therefore proportional to the vertex degree, and given by
\begin{align}\label{eq:DTRWlp}
	\bm{\pi}_j = \frac{D_{jj}}{\text{Tr}(D)}=\frac{\sum_{i}A_{ij}}{\sum_{i}\sum_{j}A_{ij}}
\end{align}

Note that in cases where a graph only has even-length closed loops, the walker will only occupy sites an even distance from the initial state at even time-steps --- causing the walker to never converge to its stationary distribution $\bm{\pi}$. In such cases, it is useful to redefine the DTRW so that at every time-step, the walker only has a $\epsilon$ probability of moving as per the transition matrix:
\begin{align}
	\mathbf{P}^{(n+1)}&=\epsilon T\mathbf{P}^{(n)}+(1-\epsilon)\mathbf{P}^{(n)}\notag\\
	&=\epsilon(T-I)\mathbf{P}^{(n)}+\mathbf{P}^{(n)}
\end{align}
This is known as the \textit{lazy random walk}, and is sufficient to break the periodicity and ensure convergence to the limiting probability distribution $\bm{\pi}$.

We can interpret each time-step of the lazy random walk as corresponding to a time of $\epsilon$ \cite{childs2010}. Thus, by rearranging this equation and taking the limit $\epsilon\rightarrow0$, 
\begin{align}
	&\lim_{\epsilon\rightarrow0}\frac{\mathbf{P}^{(n+1)}-\mathbf{P}^{(n)}}{\epsilon} = -(I-T)\mathbf{P}^{(n)}
\end{align}
we arrive at the Master equation, a stochastic Markovian process governing the time evolution of the \textit{continuous-time random walk} (CTRW):
\begin{align}
	\frac{d}{dt}\mathbf{P}(t)=-L\mathbf{P}(t)
\end{align}
with solution $\mathbf{P}(t)=e^{-Lt}\mathbf{P}(0)$, where $L$ is the normalized graph Laplacian,
\begin{align}
	L=I-T=(D-A)D^{-1}
\end{align}
such that $e^{-L t}$ is stochastic, and the walk is probability conserving. For the CTRW, the steady-state limiting probability distribution,
\begin{align}
	\bm{\pi}=\lim_{t\rightarrow\infty}e^{-L t}\mathbf{P}(0)
\end{align}
must satisfy the equation $e^{-L t}\bm{\pi}=\bm{\pi}$. After expanding the matrix exponential as a Taylor series, it can be seen that this is equivalent to $L\bm{\pi}=\bm{0}$; i.e. $\bm{\pi}$ is the nullspace of $L$. Note that since $L=I-T$, $L\bm{\pi}=(I-T)\bm{\pi}=\bm{0}~\Rightarrow~T\bm{\pi}=\bm{\pi}$, and thus the CTRW limiting distribution and the DTRW limiting distribution (\autoref{eq:DTRWlp}) are identical.

\subsection{Continuous-time quantum walks}
The CTRWs quantum analogue\footnote{We don't describe the quantum analogue of the DTRW here --- the discrete-time quantum walk (DTQW) --- as, unlike the classical case, it has a highly nontrivial relationship with the CTQW due to the existence of an additional `coinspace'. For a good introduction to the coined DTQW, see \citet{kempe2003} or \citet{wang2014}, whilst \citet{szegedy2004} offers a good introduction to the Szegedy DTQW formalism.}, the continuous-time quantum walk (CTQW) on graph $G$, has its time evolution governed instead by the Schr\"odinger equation \cite{farhi1998},
\begin{align}
	i \hbar\frac{d}{dt}\ket{\psi(t)} = H\ket{\psi(t)}
\end{align}
where $H$ is the system Hamiltonian, encoding the discrete structure of the underlying graph $G$, and $\ket{\psi(t)}=\sum_j \alpha_j(t)\ket{j}$ the complex-valued state vector. We use atomic units from hereon, and thus set $\hbar=m=e=1$. The general solution to the system is
 \begin{align}
 	\ket{\psi(t)}=U(t)\ket{\psi(0)}=e^{-iHt}\ket{\psi(0)}
 \end{align}
Consistent with standard quantum formalism, $\alpha_j(t)=\braket{j}{\psi(t)}\in\mathbb{C}$ is the probability amplitude, and $|\alpha_j(t)|^2$ the corresponding probability, of the walker being found at node $j$ after time $t$. Unlike the classical CTRW, the CTQW gains properties characteristic of quantum systems --- including time reversibility (hence, no limiting state) and superposition, allowing propagation through networks quadratically faster than its classical counterpart \cite{farhi1998,childs2003}. However, the CTQW is no longer a stochastic process, but rather deterministic; the probabilistic nature of the walk comes from measuring the quantum state, rather than the walk's dynamics \cite{whitfield2010}.

It is important to note that there are two competing conventions for the CTQW Hamiltonian that are ubiquitous in the field; the adjacency matrix ($H=A$) and the (combinatorial) Laplacian ($H=D-A$) \cite{wong2016}. Both provide similar dynamics (and are identical for degree regular graphs), with each being preferred for particular applications --- the adjacency matrix for simplicity in quantum computation calculations, and the Laplacian for its discrete approximation to the kinetic energy operator of quantum mechanics. In this study, we will be using $H=A$, for reasons that will come clear in subsequent sections.

\section{Graph centrality}
\label{sec:centrality}
In the study of network structure and graph theory, centrality measures are an integral tool, allowing determination and ranking of vertices deemed to be most important. Due to the large number of physical systems that can be modelled as networks, this has seen wide application across multiple disciplinary fields, including technology (ranking web sites for search engines \cite{brin1998}, power distribution \cite{knoke1983}), business (organisational management \cite{brass1984,ibarra1993,ibarra1993-a}), biology (grooming networks in macaques \cite{sade1989}), and biochemistry (finding active sites in proteins \cite{amitai2004}).

At its most basic, a graph centrality measure $C$ satisfies the following properties:
\begin{itemize}
\item $C:G(V,E)\rightarrow \mathbb{R}^{|V|}$ is a function or algorithm that accepts a graph, and returns a real-valued vector over the set of vertices $V$.
\item Higher values are provided to vertices deemed more `important' or `central' to the graph structure, with lower values provided to vertices with a reduced `importance' or `centrality'.
\end{itemize}
However, what constitutes `importance' is subjective --- it depends on the application or model to be analysed, and how information `flows' throughout the network \cite{borgatti2005}. For example, information might flow predominantly though paths (a sequence of unique edges and vertices --- characteristic of bacterial and viral infections \cite{rothenberg1995}), trails (vertices can be revisited but each edge is only traversed once --- the flow of gossip in social networks \cite{grosser2010}), and walks (where there is no restrictions on edge and vertex sequences --- for example bank note exchange in a population). Moreover, this flow can occur through serial duplication (travelling via one edge at each timestep --- gift exchange) or parallel duplication (traversing multiple edges simultaneously --- radio broadcasting). 

Thus, it is important to apply a centrality measure that models information flow corresponding to the network under study; failure to do so may result in poor results, and even the inability to correctly interpret the results \cite{borgatti2005}. To deal with this plethora of scenarios, various classical centrality measures have been introduced: degree centrality, eigenvector centrality, betweeness centrality, closeness centrality, and PageRank, among others. Of these, degree, eigenvector and PageRank centrality are what is known as \textit{radial parallel duplication} measures, which measure network flow via walks emanating from or terminating at particular nodes \cite{borgatti2006}.

\subsection{Degree centrality}

The degree centrality measure, calculated via the row-sums of the adjacency matrix,
\begin{align}
C^{(deg)}_j=\frac{\text{deg}(v_j)}{\sum_{k}\text{deg}(v_k)}=\frac{\sum_{i}A_{ij}}{\sum_{i}\sum_{j}A_{ij}}
\end{align}
is based on walks of length one emanating from each vertex, and is useful in cases when dealing with direct and immediate influence between nodes. Further, it can be seen that the limiting probability distribution of classical random walks are proportional to the node degree, allowing the degree centrality to be simulated via a Markovian process.

\subsection{Eigenvector centrality}

Eigenvector centrality, on the other hand, is given by $C^{(ev)}_j=\mathbf{v}_j$, where $\mathbf{v}$ is an eigenvector of the adjacency matrix $A\mathbf{v}=\lambda\mathbf{v}$ corresponding to the maximum eigenvalue to ensure, via the Perron-Frobenius theorem, that the ranking remains strictly positive. It has been shown by \citet{bonacich1987} that the eigenvector centrality is proportional to the row-sums of matrix $S$, $\mathbf{v}_j\propto\sum_{i}S_{ij}$, where
\begin{align}
	S=A+\frac{1}{\lambda}A^2+\frac{1}{\lambda^2}A^3+\cdots = \sum_{n=1}^\infty\lambda^{1-n}A^n
\end{align}
i.e. the eigenvector centrality counts walks of \textit{all} lengths, weighted inversely by length, from each node. Thus, unlike the degree centrality, the eigenvector centrality considers long-term `indirect' influence --- if a vertex is connected to another node with a high number of connections, the first vertex will likewise have a high centrality measure. Consequently, rather than model the eigenvector centrality via the DTRW --- which may only sample adjacent vertices at each time-step --- we can instead use the continuous-time random walk (CTRW). Due to its matrix exponential time-propagator, the CTRW performs walks of \textit{all} lengths at each infinitesimal time-step $dt$.

\subsection{PageRank}

One final classical centrality measure which necessitates introduction is the Google PageRank \cite{brin1998}. A variation of the eigenvector centrality, PageRank was developed as a ranking algorithm for sites on the world wide web, and has accumulated significant prestige as the algorithm behind the Google search engine. In this context, vertices represent websites, with directed edges the links between them. Due to the need to take into account direction, issues arise with eigenvector centrality --- namely, nodes with in-degree but no out-degree (`dangling nodes') accumulate probability, due to the adjacency matrix being non-stochastic. To address this issue, the eigenvector centrality method is instead applied to the Google matrix $G$,
\begin{align}
	G = \alpha E+\frac{1}{N}(1-\alpha)J, ~~~0\leq\alpha\leq 1
\end{align}
where $N$ is the number of vertices in the graph, $E$ is the \textit{patched adjacency matrix}, column-normalised to ensure $G$ is stochastic,
\begin{align}
	E_{ij}=\begin{cases}
	A_{ij}/\sum_k A_{kj}, & \sum_k A_{kj}\neq0\\
	1/N, &\sum_k A_{kj}=0
	\end{cases}
\end{align}
and $J$ the all one's matrix. The addition of $J$ is to provide a small `random surfer effect', i.e. a non-zero uniform probability that a walker at a particular vertex can jump to \textit{any} other vertex, even in cases of non-adjacency. In practise, $\alpha$ is generally chosen to be $0.85$, providing a good compromise between information flow via hyperlinks and the random surfer effect.

Once the Google matrix is calculated, the PageRank centrality measure is then applied by solving the eigenvector equation
\begin{align}\label{eq:PReqn}
	G\bm{x}=\bm{x}
\end{align}
as, per the Perron-Frobenius theorem, the eigenvector corresponding to the largest eigenvalue ($\lambda=1$ for PageRank, as $G$ is stochastic) will be strictly positive. Note that this equation is identical to that of a DTRW; thus, the PageRank can be modelled as a DTRW with $G$ taken to be the transition matrix. Nevertheless, when $\alpha<1$, the PageRank continues to model its centrality measure on walks of all lengths, due to the random surfer effect. To see this explicitly, it can be easily shown that in the case $0\leq\alpha<1$,
\begin{align}
	&G\bm{x}=\bm{x}\notag\\
	&\Rightarrow~~\alpha E\bm{x}+\frac{1}{N}(1-\alpha)J\bm{x}=\bm{x}\notag\\
	&\Rightarrow~~(I-\alpha E)\bm{x}=\frac{1}{N}(1-\alpha)J\bm{x}
\end{align}
 has the exact solution
\begin{align}
	\bm{x}&=\left(I-\alpha E\right)^{-1}\frac{1}{N}(1-\alpha)J\bm{x}\notag\\
	&=(1-\alpha)\left(\sum_{k=0}^\infty\alpha^kE^k\right)\sum_{j=1}^N\frac{\mathbf{e}_j}{N}\notag\\
	&\hspace{-1cm}\therefore~~\bm{x}_i=\frac{1}{N}(1-\alpha)\sum_{j=1}^N\left(\sum_{k=0}^\infty\alpha^kE^k\right)_{ij}
\end{align}
and therefore $\bm{x}$ is calculated using walks of $k$ lengths for all $k\in\mathbb{N}$, weighted by $\alpha^k$ \cite{gleich2016}.

\subsection{Random walk centrality}
The random walk centrality, unlike the centralities previously discussed, is not a radial volume based measure (counting the number of walks between each pair of nodes) but rather a radial \textit{length} based measure, quantifying the length of the walks between nodes \cite{borgatti2006}. Alternatively, this can be interpreted as a measure of the expected time for information to arrive at a particular node; i.e., the effectiveness or speed of communication \cite{borgatti2005}. The Random Walk Centrality (RWC) measure, introduced by \citet{noh2004} and based on a DTRW, is given by 
\begin{align}
	C^{(RWC)}_j = \frac{\bm{\pi}_j}{\tau_j}
\end{align}
where $\bm{\pi}$ is the random walk limiting distribution, and
\begin{align}
	\tau_j=\sum_{n=0}^{\infty}\left(T^n_{jj}-\bm{\pi}_j\right)
\end{align}
is the characteristic relaxation time of vertex $j$.


\subsection{Quantum centrality measures}
The above described walk-based centrality measures have been classical in nature. However, in recent years several quantum centrality measures have been proposed --- ranging from quantizations of the aforementioned classical measures to wholly new proposals --- that take advantage of the exponential speedups offered by quantum computation. For example, the Quantum PageRank, (introduced by \citet{paparo2012}\cite{paparo2013} and extended by \citet{loke2017}) utilizes the Szegedy quantum walk \cite{szegedy2004} (a DTQW formulation) to quantize the directed Markov chains encoded by the Google matrix, before taking the long-time average of the walks probability distribution; in essence, providing a quantum analogue of PageRank centrality.

The quantum stochastic walk (QSW) is another approach, which makes use of the Lindblad master equation to introduce environmental decoherence to a CTQW \cite{loke2017,sinayskiy2013,falloon2016}. In practice, this has the effect of creating a continuous-time walk continuum parametrized by $\omega$, with $\omega=0$ (no dephasing) corresponding to a purely quantum walk (CTQW) on an undirected graph, and $\omega=1$ (complete dephasing) corresponding to a purely classical walk (CTRW) over a digraph. By restricting the domain to $0<\omega\ll 1$, quantum dynamics and the resulting quantum speedup is retained, however the walker will eventually converge to the CTRW limiting probability distribution \cite{whitfield2010}. Similar to the Quantum PageRank, the centrality measure is then given by the long-time average of the probability distribution.

Finally, \citet{Berry2010} proposed a novel method, in which the quantum search algorithm is applied to graph structures via DTQW --- the resulting frequency of successful search probability was shown to correlate with the (lazy) Random Walk Centrality of \citet{noh2004}.
Thus, the quantum centrality scheme of \citet{Berry2010} differs from the two previous quantum centrality schemes, as it considers the mean speed of the walker in transmitting information over the network --- it is a form of \textit{quantum closeness centrality}. 

Unfortunately, when it comes to physically implementing these quantum centrality measures, we run into various issues. In all three cases, due to the use of either a coin state (DTQW) or an environment (QSW), the size of the statespace must be significantly increased, taking us beyond the experimental capability to simulate quantum graph centrality of even simple graph structures. For example, for a graph of $N$ vertices, the Szegedy DTQW formulation used in the Quantum PageRank scheme requires a statespace of size $N^2$. As such, the ability to physically realise these quantum centrality measures is currently beyond our reach.

Instead, in the following section we propose a CTQW-based centrality measure --- building on the foundation of classical radial centrality measures such as eigenvector centrality, whilst allowing us to take advantage of the quantum speedup afforded over the CTRW, and utilizing a significantly smaller statespace than the QSW and DTQW.

\section{CTQW-based centrality measure}
\label{sec:ctqwcentrality}

Similarly to the Quantum PageRank and the QSW centrality measures, as the time-evolution of the CTQW is determined by the Hamiltonian --- and thus the underlying network structure --- one method for extracting the centrality information is to simply start the walker in an equal superposition of all vertex states, $\ket{\psi(0)}=\frac{1}{\sqrt{N}}\sum_j \ket{j}$ (so as not to bias any one particular vertex), and compare the time-average probability of locating the walker at each vertex.

Now, convention allows for two choices for the Hamiltonian --- we may choose either the adjacency matrix $A$ or the graph Laplacian $L$ (given by $L_{ij}=\delta_{ij}\sum_{k}A_{ik}-A_{ij}$, a discrete approximation to the continuous-space Laplacian). However, the construction of the Laplacian ensures that equal superposition state is always an eigenvector, resulting in a stationary time-evolution:
\begin{align}
	U\ket{\psi(0)}=e^{-i L t}\left(\frac{1}{N}\sum_j \ket{j}\right) = \frac{1}{N}\sum_j \ket{j}~~\forall t
\end{align}
As such, the Laplacian is ill-suited for a CTQW centrality measure, as it will be unable to distinguish vertices more central to the network structure. This is not the case of the adjacency matrix; thus, for the remainder of this work, we will set $H=A$.

To briefly summarise, the proposed CTQW centrality scheme works as follows:
\begin{enumerate}
\item Prepare the quantum walker in an initial equal superposition over all vertex states: $\ket{\psi(0)}=\frac{1}{N}\sum_j \ket{j}$
\item Propagate the walker for time $t\gg 0$: $\ket{\psi(t)}=e^{-iHt}\ket{\psi(0)}$, where $H=A$ is the graph adjacency matrix.
\item Calculate the long-time average probability distribution of finding the walker at each vertex: 
\begin{align}
C^{(CTQW)}_j= \lim_{\tau\rightarrow\infty}\frac{1}{\tau}\int_0^{\tau} |\braket{j}{\psi(t)}|^2~dt
\end{align}
\end{enumerate}

To fully ascertain the reliability of the proposed CTQW centrality measure, we will consider both a simple example (allowing us to qualitatively assess the measures performance), as well as a rigorous statistical analysis comparing the CTQW measure to PageRank over an ensemble of randomly generated graphs. \citet{freeman1978}, in his discussion of the canonical formulations of centrality measures, noted that degree, closeness and betweeness centralities all attain their highest values for the central node of the star graph; \citet{borgatti2006}, in reviewing Freeman's work, suggested that this may serve as a defining characteristic of a `proper' centrality measure. Thus, let us consider a 4-vertex star graph as an example of the proposed CTQW centrality measure.

\begin{figure}
	\centering
	\includegraphics[scale=0.6]{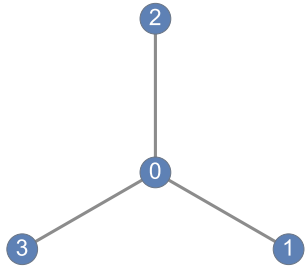}
	\caption{4-vertex star graph}
	\label{fig:stargraph}
\end{figure}
\begin{figure}
	\centering
	\includegraphics[scale=0.7]{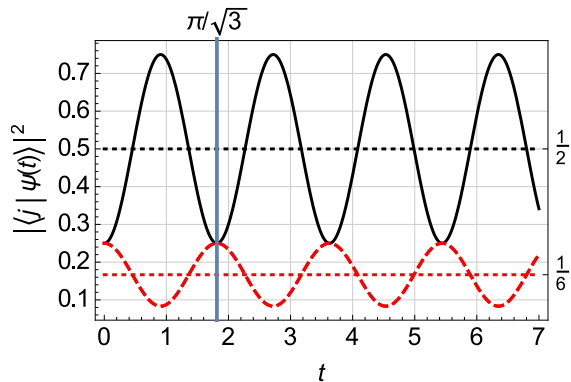}
	\caption{CTQW probability at vertex 0 (black, solid) and vertex 1,2,3 (red, dashed) on a 4-vertex star graph. The initial state is a equal superposition of all vertex states. The dotted lines show the respective long-time averaged probability of the respective vertices, with the blue vertical line denoting one period ($T=\pi/\sqrt{3}$).}
	\label{fig:sg4prob}
\end{figure}

For the 4-vertex star graph shown in \autoref{fig:stargraph}, the adjacency matrix is 
\begin{align}
	A= \left[\begin{matrix}
		0 & 1 & 1 & 1\\
		1 & 0 & 0 & 0\\
		1 & 0 & 0 & 0 \\
		1 & 0 & 0 & 0
	\end{matrix}\right]
\end{align}
with the first vertex (vertex 0) the central node. In this case, the time-evolution operator is given by
\begin{align}\label{eq:unitary}
	U(t)= \frac{1}{3}\left[\begin{matrix}
		3c(t) & s(t) & s(t) & s(t)\\
		s(t) & c(t)+2 & c(t)-1 & c(t)-1\\
		s(t) & c(t)-1 & c(t)+2 & c(t)-1 \\
		s(t) & c(t)-1 & c(t)-1 & c(t)+1
	\end{matrix}\right]
\end{align}
where $c(t)=\cos(\sqrt{3}t)$ and $s(t)=-i\sqrt{3}\sin(\sqrt{3}t)$. Using this operator to propagate from an initial equal superposition of vertex states $\ket{\psi(0)}=\frac{1}{4}\sum_j \ket{j}$, the probability of locating the walker on vertex $j$ at time $t$ is
\begin{align}\label{eq:probability}
	&|\braket{j}{\psi(t)}|^2=|\braket{j}{U(t)|\psi(0)}|^2\notag\\
	&~~~~~ =\left[\frac{1}{2}-\frac{1}{4}c(2t)\right]\delta_{j0}+\left[\frac{1}{6}+\frac{1}{12}c(2t)\right]\sum_{j'=1}^3\delta{jj'}
\end{align}
(\autoref{fig:sg4prob}). Noting that this probability distribution is periodic with period $T= \pi/\sqrt{3}$, the CTQW centrality measure becomes
\begin{align}
	C^{(CTQW)}_j &= \lim_{\tau\rightarrow\infty}\frac{1}{\tau}\int_{0}^{\tau}|\braket{j}{\psi(t)}|^2~dt\notag\\
	& = \frac{1}{T}\int_{0}^{T}|\braket{j}{\psi(t)}|^2~dt
\end{align}
yielding values of $1/2$ for $j=0$, and $1/6$ for $j=1,2,3$. This fits well with what would be expected intuitively --- the central vertex (vertex 0) has the highest time-averaged probability, indicating a high centrality measure, whilst the remaining vertices (1,2,3) are equivalent and have an equal and lower ranking. The proposed CTQW centrality measure therefore satisfies one of the defining properties of centrality measures; however, a detailed statistical analysis is required to properly assess its behaviour on general graphs.

\section{Statistical analysis}
\label{sec:stats}

To investigate the reliability of this newly proposed quantum centrality algorithm, it is pertinent to compare its ranking results to classical algorithms on large random graphs. To do so, we consider two classes of random graphs --- Erd\H{o}s-R\'enyi networks, and scale-free networks. A random Erd\H{o}s-R\'enyi graph, denoted $G(N,p)$, is comprised of $N$ vertices with edges randomly distributed as per the Bernoulli distribution, with probability $p$ \cite{erdoes1959,erds1960}. For such a network, the vertex degree distribution $P(k)$ (the fraction of vertices with degree $k$) is binomial in form, resulting in most vertices with degree close to $np$, the mean number of connections. Scale-free networks, on the other hand, are characterised by a power law vertex distribution $P(k)\sim k^{-\gamma}$, due to the presence of a small number of highly connected `hubs' --- with a majority of vertices exhibiting significantly lower degree \cite{barabasi1999,barabasi2000}. As such, they form an important tool in modelling real-life networks with similar characteristics, such as social networks, the World Wide Web, and biochemical molecules \cite{albert2002,song2005}.

\subsection{Correlation to classical measures}
\begin{figure}
	\centering
	\subfigure{\includegraphics[scale=0.7]{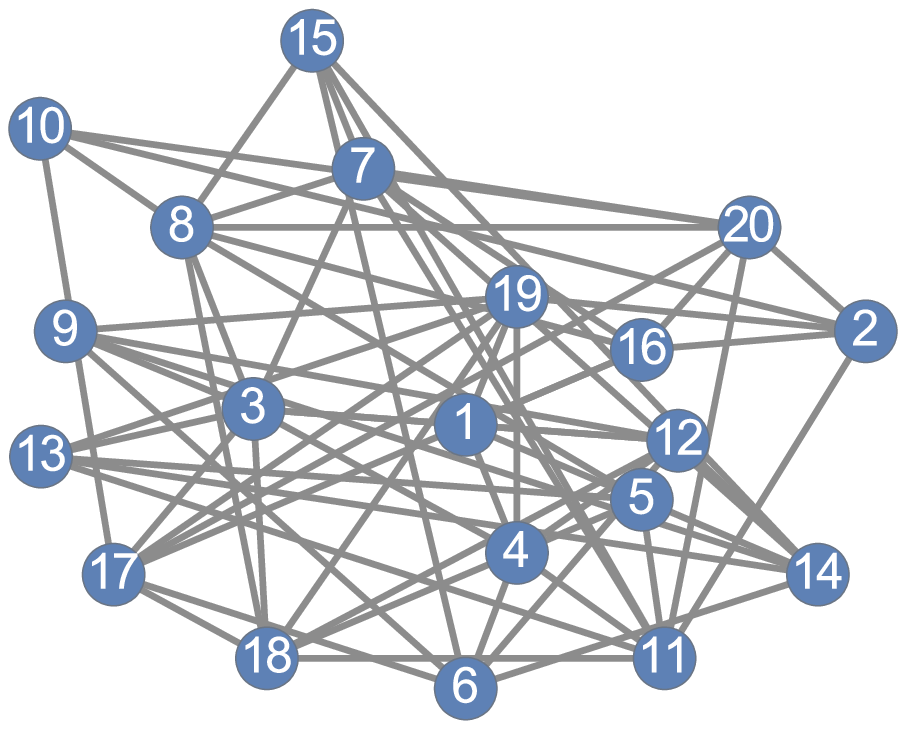}}
 	\subfigure{\includegraphics[scale=0.7]{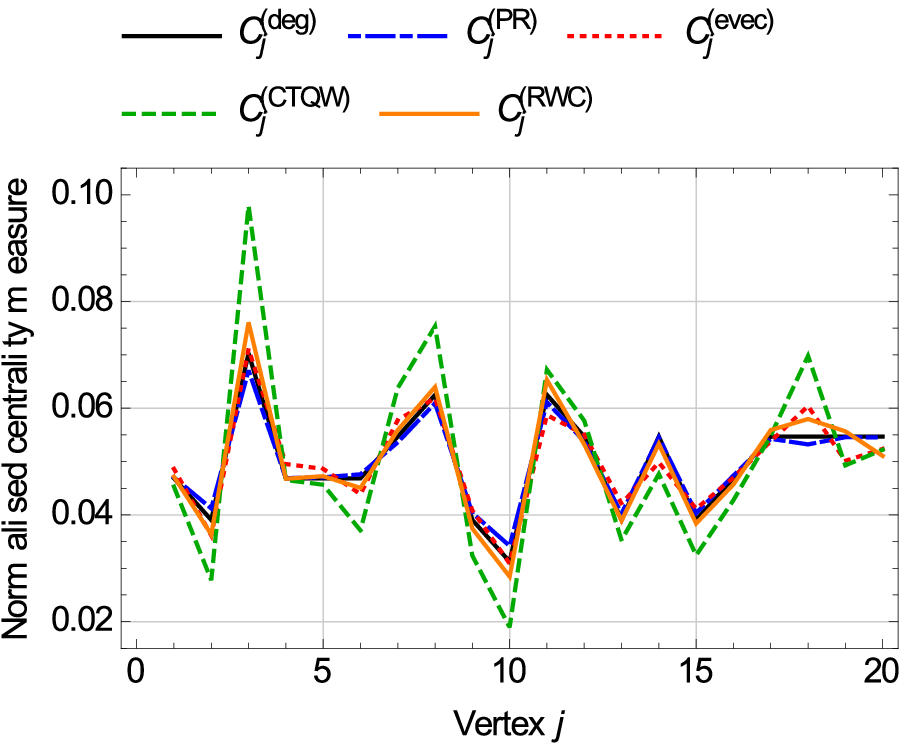}}
	\caption{\textbf{Left:} Randomly generated Erd\H{o}s-R\'enyi graph $G(20,0.3)$ \textbf{Right:} Normalised vertex centrality values for vertex $j$ on $G(20,3)$. Measures shown are degree centrality, PageRank, eigenvector centrality, CTQW centrality, and RWC centrality. 
	}
	\label{fig:corr}
\end{figure}
 Firstly, let's investigate correlation between the CTQW centrality measure and classical measures. As centrality measures only provide useful information for the top 5 or so valued vertices (with noise growing successively larger for lower ranked nodes \cite{lawyer2015}), we will consider a randomly generated 20-vertex Erd\H{o}s-R\'enyi graph $G(20,0.3)$, and use this as the basis of our correlation test. The graph generated and its respective vertex centrality values (calculated using degree centrality, PageRank, eigenvector centrality, CTQW centrality, and RWC centrality) are shown in \autoref{fig:corr}. Qualitatively, it can be seen that all centrality measures strongly agree on the top-ranked vertices, with slight variations for the lower ranked vertices, as is expected.
\begin{table}
\begin{tabular}{r|ccccc}
	&Degree & PageRank & Eigenvector & CTQW & RWC\\
	\hline
	Deg & 1. & 0.316 & 0.232 & 0.232 & 0.411 \\
	PR & 0.316 & 1. & 0.011 & 0.011 & 0.189 \\
	EVec & 0.232 & 0.011 & 1. & 1. & 0.568 \\
	CTQW& 0.232 & 0.011 & 1. & 1. & 0.568 \\
	RWC & 0.411 & 0.189 & 0.568 & 0.568 & 1. 
\end{tabular}
\vspace{7pt}

\begin{tabular}{r|ccccc}
  & \text{Degree} & \text{PageRank} & \text{Eigenvector} & \text{CTQW} & \text{RWC} \\
  \hline
 \text{Deg}~ & 1. & 0.276 & 0.163 & 0.15 & 0.236 \\
 \text{PR}~ & 0.276 & 1. & 0.107 & 0.076 & 0.12 \\
 \text{EVec} ~& 0.163 & 0.107 & 1. & 0.592 & 0.277 \\
 \text{CTQW} ~& 0.15 & 0.076 & 0.592 & 1. & 0.228 \\
 \text{RWC} ~& 0.236 & 0.12 & 0.277 & 0.228 & 1.
\end{tabular}
\caption{Kendall's tau coefficient comparing the vertex rankings for the labelled centrality measures, calculated for the Erd\H{o}s-R\'enyi graph shown in \autoref{fig:corr} (left) and averaged over an ensemble of 100 random Erd\H{o}s-R\'enyi graphs $G(20,0.3)$ (above).}
\label{tab:kendalltau}
\end{table}

To get a more quantitative understanding of the correlation, we employ Kendall's tau rank correlation coefficient \cite{kendall1938} ($\tau\in[-1,1]$, where $\tau=1$ indicates perfect correlation between ranked lists, $\tau=0$ indicates no correlation, and $\tau=-1$ indicates perfect anti-correlation). Kendall's tau correlation coefficients for \autoref{fig:corr} are shown in \autoref{tab:kendalltau}; for additional robustness, this analysis is repeated and averaged over an ensemble of 100 randomly generated $G(20,0.3)$ graphs. It can be seen that there is a significant correlation between the CTQW centrality ranking and the eigenvector centrality ranking ($\tau=0.592$ averaged across the ensemble). If we recall that the CTQW propagator is the matrix exponential, this is perhaps not so surprising --- the CTQW centrality scheme appears to be ranking the graph vertices in a similar fashion to the eigenvector centrality, by considering walks of \textit{all} lengths emanating from each vertex weighted inversely by length.

Note that Kendall's tau coefficient tells us how correlated the \textit{entire} ranked lists are, allowing us to classify the centrality measures based on how they encode information flow through the network. However, beyond the topmost ranked vertices, centrality measures convey very little useful information regarding remaining vertices --- this is more the domain of \textit{influence measures} \cite{lawyer2015,liu2013}. As such, Kendall's tau coefficient is not useful for determining \textit{general agreement} between centrality measures on the location of the \textit{most central} nodes. For example, in \autoref{fig:corr} it can be seen that the PageRank (a radial volume measure) and random walk centrality (a radial length measure) are in total agreement on the location of the top 3 most central vertices, whilst exhibiting low correlation ($\tau=0.189$).

\subsection{Agreement on top-ranked vertices}

Here, we consider ensembles of larger Erd\H{o}s-R\'enyi and scale-free graphs, and compare the CTQW centrality to the eigenvector centrality (its closest classical analogue), and to the PageRank (the classical centrality measure with arguably the most impact in the last decade). This analyses will allow us to verify the behaviour of the CTQW centrality for large graphs of varying degree distributions.

\begin{figure}
	\centering
	\subfigure[The average CTQW centrality measure (black) compared to the average PageRank measure (red, dashed)]{\includegraphics[scale=0.5]{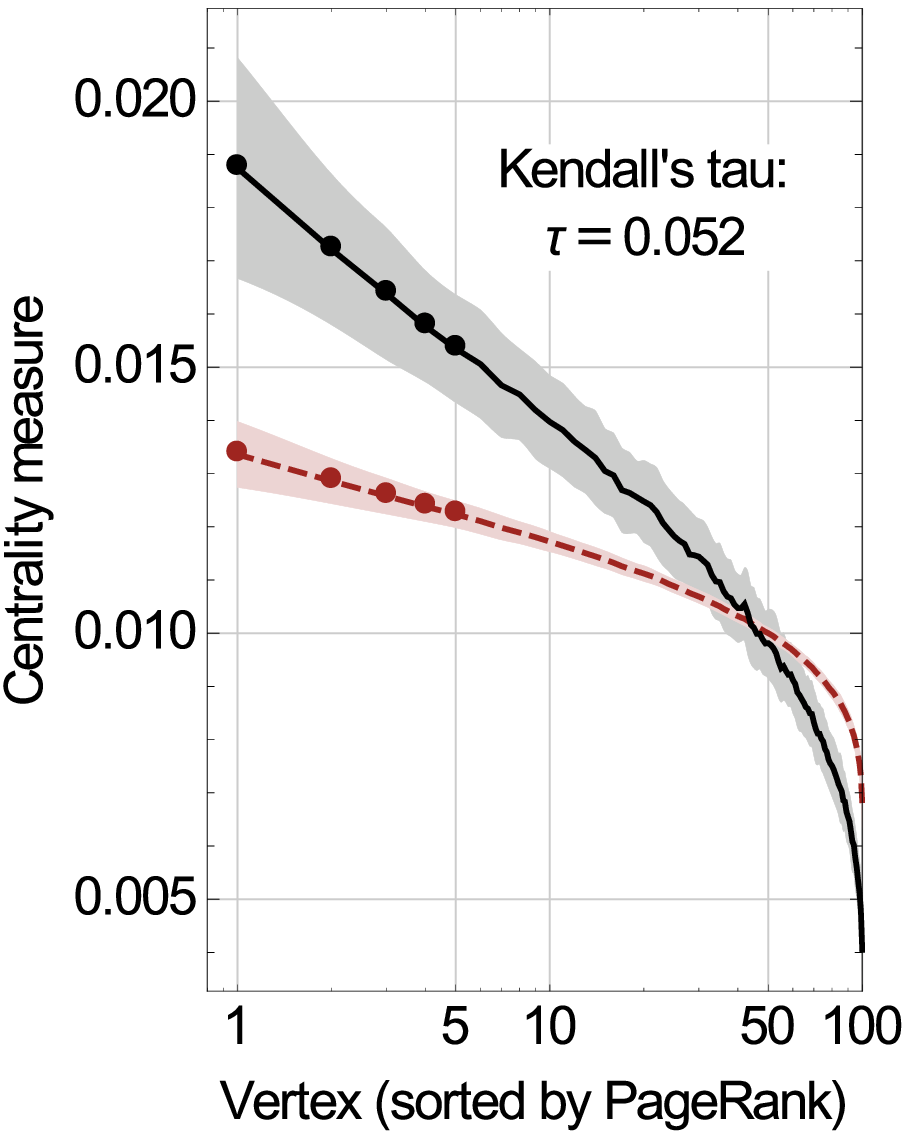}\hspace{-10pt}\includegraphics[scale=0.5]{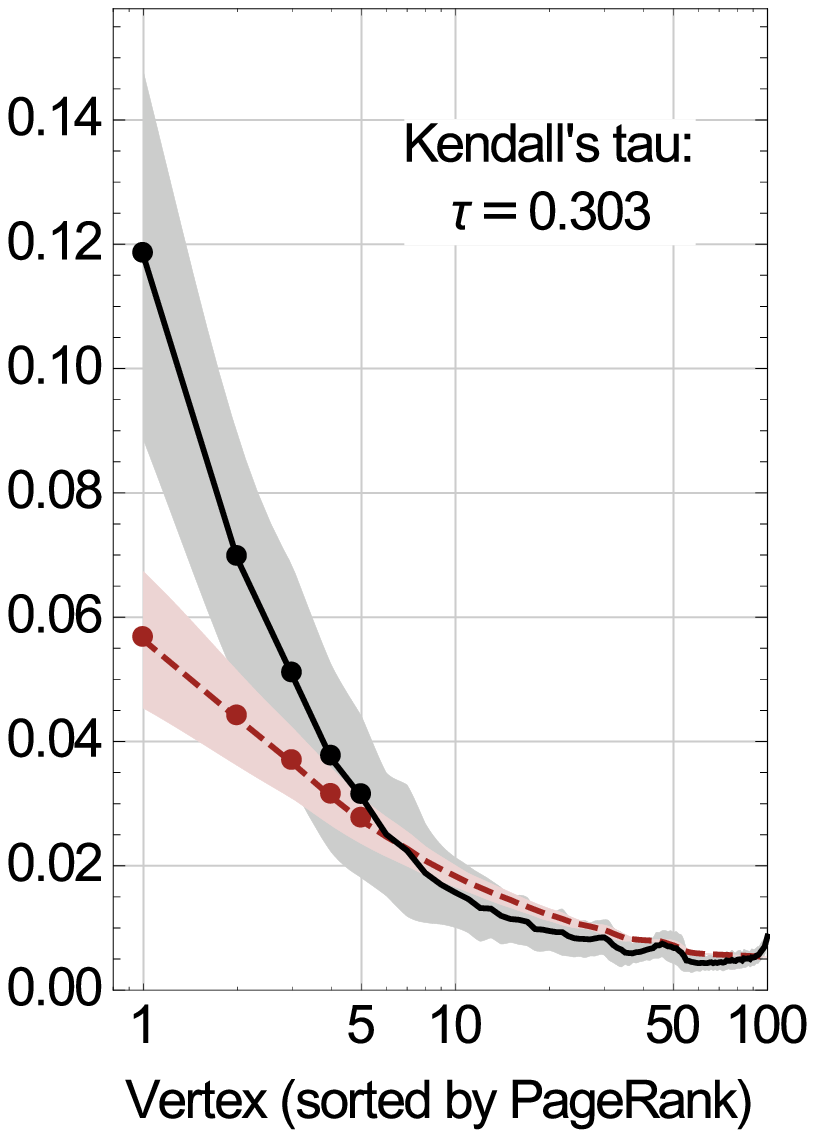}}
	\subfigure[The average CTQW centrality measure (black) compared to the average eigenvector centrality measure (blue, dashed)]{\includegraphics[scale=0.5]{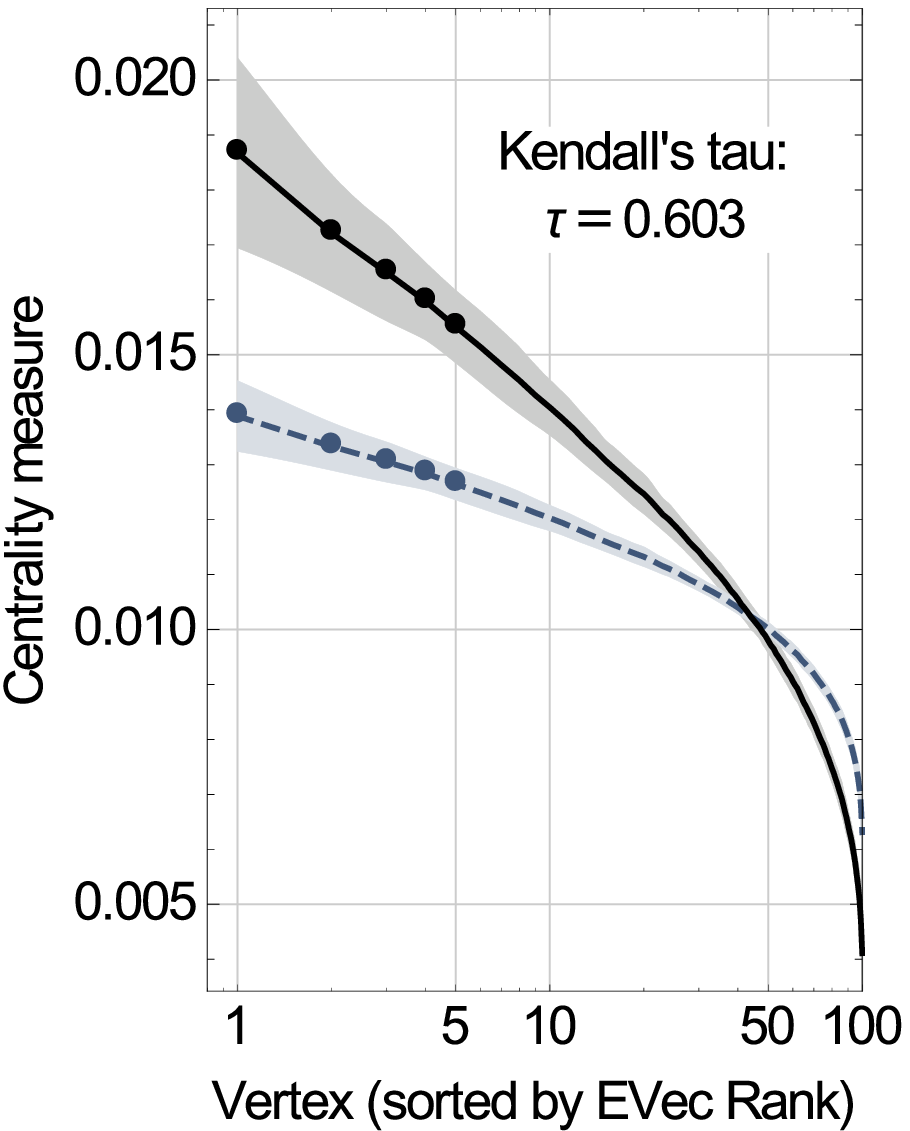}\hspace{-10pt}\includegraphics[scale=0.5]{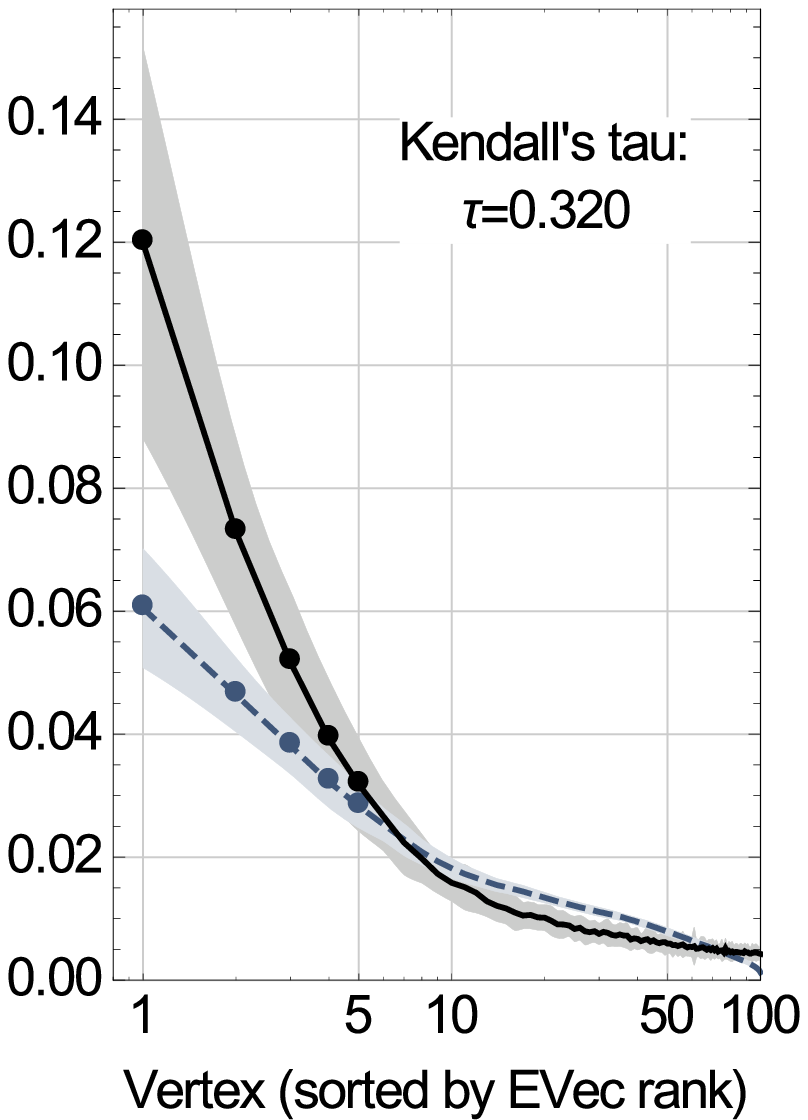}}
	\caption{The average CTQW centrality measure compared to classical centrality measures for vertices in an ensemble of 200 Erd\H{o}s-R\'enyi graphs ({left}) and 200 scale-free graphs ({right}). The Erd\H{o}s-R\'enyi graphs have parameters $N=100$, $p=0.3$. The scale-free networks are constructed via the Barab\'asi-Albert algorithm with $N=100$ and $m=2$ edges added at every generation. The shaded areas represent one standard deviation from the average centralities, and the top 5 ranked vertices are shown by the markers. }
	\label{fig:ERSFplot}
\end{figure}
We begin by generating an ensemble of 200 Erd\H{o}s-R\'enyi and scale-free graphs (the latter by way of the Barab\'asi-Albert algorithm), and calculating the average PageRank, eigenvector, and CTQW centrality measures over the ensemble. These results are shown in \autoref{fig:ERSFplot}. It can be seen that, on average, the CTQW ranking agrees with the classical algorithms on the location of the five most central vertices, whilst also following the following the same general trendline (binomial for the Erd\H{o}s-R\'enyi, power law for the scale-free). In fact, the CTQW measure for the top 5 vertices \textit{outperforms} that of the PageRank and eigenvector centrality, by assigning a higher centrality measure, perhaps allowing for greater distinguishability when sampled experimentally. However, it appears that this comes at the cost of larger measure variance compared to the PageRank.

\begin{figure}
	\centering
	\subfigure{\includegraphics[scale=0.95]{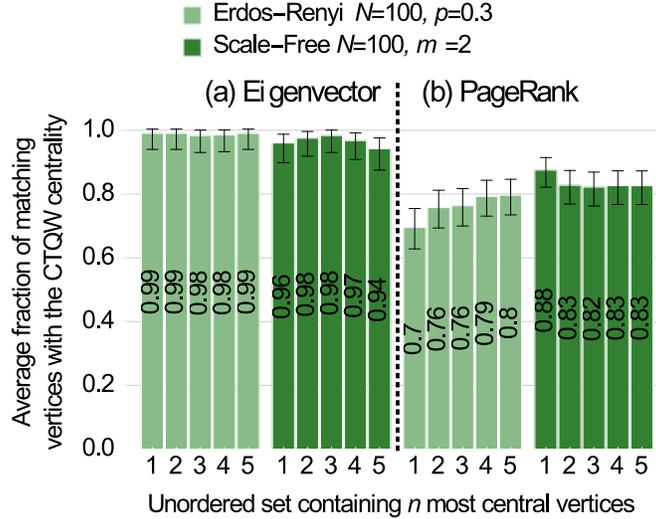}}
	\caption{Charts showing the agreement between the CTQW centrality ranking and the (a) PageRank and (b) eigenvector centrality ranking for an ensemble of 100 Erd\H{o}s-R\'enyi and 100 scale-free graphs. Each bar represents the unordered set containing the $n$ most central vertices as determined by the PageRank and CTQW measures, whilst the vertical axis gives the average fraction of matching vertices between the two sets. The error bars indicate the Agresti-Coull 95\% confidence interval.}
	\label{fig:bc}
\end{figure}
We now attempt to quantify the `agreement' regarding the top-ranked vertices between the quantum and classical measures. Previously Kendall's tau ranked-list coefficient was used; whilst this works great in determining correlations between various ranked lists, it is less useful in quantifying how often the ranked lists agree on their first few values. Thus, we detail an alternative approach. For each graph in the ensemble, unordered sets containing the $n$ most central vertices according to each measure was compared --- the fraction of matching vertices providing a quantitative value for the agreement between the two measures (termed the \textit{agreement factor}). These agreement factors were then averaged over the entire ensemble, with uncertainty approximated by calculating the Agresti-Coull 95\% confidence interval --- the results are presented in \autoref{fig:bc}.

It can be seen that the eigenvector and CTQW centrality measures are in near perfect agreement --- agreeing on the most central node 99\% of the time for Erd\H{o}s-R\'enyi networks, and 96\% of the time for scale-free networks. This is likely due to the strong correlation between the CTQW and eigenvector centralities noted previously, and indicates that this strong correlation continues to hold for larger graphs of varying degree distributions.

Turning our attention to the PageRank, we find a strong agreement with the CTQW measure, albeit not as strong as the eigenvector centrality; on the location of the most central vertex, they named the same vertex 88\% of the time for scale-free graphs, dropping to 70\% for Erd\H{o}s-R\'enyi graphs. As the number of vertices compared increases, the agreement factors decrease slightly for the scale-free and increase slightly for the Erd\H{o}s-R\'enyi, before both ending around 80\% by the time the top 5 vertices are compared. This discrepancy might be partially explained by considering the CTQW measure variance in \autoref{fig:ERSFplot}:
\begin{itemize}
\item For the Erd\H{o}s-R\'enyi graphs, a majority of vertices have degree close to the mean, leading to the top-ranked vertices having similar centrality measures. The average CTQW centrality measure of the second and third-ranked vertices lies \textit{within} the uncertainty region of the most central vertex; so even as the top 5 are easily distinguished, changes in their initial ordering might appear. 
\item For the scale-free graphs, with a small number of connected hubs, the hubs are easily distinguished by both measures. However, beyond the hubs, most vertices have similar degree due to the power law distribution --- leading to small discrepancies between the measures as more vertices are ranked.
\end{itemize}
Nonetheless, our results here show that the CTQW measure proposed works excellently as a centrality measure --- it assigns higher values to the central node of a star graph and equal lower values to the surrounding nodes, correlates well with the classical eigenvector centrality (allowing us to posit that the CTQW measure extracts centrality in a similar fashion to the eigenvector centrality, namely via weighted walks of all lengths), and generalises to arbitrary random scale-free and Erd\H{o}s-R\'enyi graphs. Thus, the proposed quantum scheme sufficiently determines node centrality, and in contrast to the Quantum PageRank algorithm (which requires computation of the dense Google matrix), preserves the sparse structure of the network in the Hamiltonian; a property that allows for known efficient quantum implementation \cite{berry2015}.

In the following section, we build off this result to experimentally implement the CTQW centrality scheme on a star graph using linear optics --- a proof-of-concept experiment and the first physical implementation (to our knowledge) of quantum centrality.

\section{Experimental realization}
\label{sec:experiment}

Linear optics enables the efficient implementation of arbitrary unitary transformation on various degrees of freedoms of single photons. For example, any $2\times 2$ unitary transformations on the polarizations of single photons can be realized by a set of half-wave plates (HWPs) and quarter-wave plates (QWPs) \cite{reck1994}. Here we aim to devise a linear-optics realization of $4\times 4$ unitary transformation using spatial and polarization degrees of freedoms of single photons.

In this experiment, we first prepare a four-dimensional equal superposition quantum state $\ket{\psi(0)}=\frac{1}{2}\sum_{j=0}^3\ket{j}$, and then perform a $4\times 4$ unitary transformations on the state. We obtain the probabilities distribution through projective measurement on the state. The unitary transformations applied on the initial state $\ket{\psi(0)}$ are $U(k\Delta t)$, where $k\in\{1,2,\dots,8\}$.

An arbitrary $4\times4$ unitary transformations can be decomposed using the cosine-sine decomposition method~\cite{stewart1977,stewart1982,paige1994,sutton2008,dhand2015,goyal2015}. For each unitary transformation in $U(k\Delta t)$, there exist unitary matrices, $\mathbb{L}$, $\mathbb{S}$, $\mathbb{R}$, such that $U=\mathbb{L}\mathbb{S}\mathbb{R}$
where $\mathbb{L}$ and $\mathbb{R}$ are block-diagonal
\begin{align}
  \mathbb{L}=\left[\begin{array}{c|c}
    L & 0\\
    \hline
    0 & L^\prime\\
  \end{array}\right], \mathbb{R}=\left[\begin{array}{c|c}
    R & 0\\
    \hline
    0 & R^\prime\\
  \end{array}\right],
\end{align}
and $\mathbb{S}$ is an orthogonal cosine-sine matrix:
\begin{align}
  \mathbb{S}=\left[\begin{array}{cc|cc}
  \cos\theta & 0 & \sin\theta & 0\\
  0 & 1 & 0 & 0\\ \hline
   -\sin\theta & 0 & \cos\theta & 0\\
   0 & 0 & 0 & 1 \end{array}\right],
\end{align}
where $L$, $L'$, $R$ and $R'$ are arbitrary $2\times2$ unitary transformations on two modes. This matrix $\mathbb{S}$ can be further decomposed by a $2\times2$ unitary transformation $
S=\begin{bmatrix}
\cos\theta & \sin\theta\\
-\sin\theta & \cos\theta\\
\end{bmatrix}$ on the subspace spanned by the modes $\{\ket{0},\ket{2}\}$ and $\mathbb{I}=\begin{bmatrix}
1 & 0\\
0 & 1\\
\end{bmatrix}$ on the subspace spanned by the modes $\{\ket{1},\ket{3}\}$.

This decomposition method can be used to decompose any higher dimensional unitary operations into series of two dimensional unitary operations, and thus our technology can be used to realised, in principle, any dimensional unitary operations.
However, it is worth noting the numbers of BDs used to prepare a 2d-dimensional state and to realise the 2-dimensional unitary operation are $d-1$ and $2^d-2$ respectively. In other words, the number of optical elements grows exponentially with the dimension of the unitary operation, and decoherence in cascaded interferometers also grows.

\begin{figure}
  \includegraphics[width=0.35\textwidth]{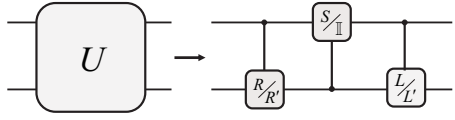}\\
  \caption{The quantum circuit for implementing the $4\times4$ unitary transformation $U$ on a two-qubit system.}
  \label{fig:circuit}
\end{figure}

For convenience, we encode the four-dimensional quantum states by two-qubit state as $\{\ket{0}=\ket{\tilde{0}\tilde{0}},\ket{1}=\ket{\tilde{0}\tilde{1}},\ket{2}=\ket{\tilde{1}\tilde{0}},\ket{3}=\ket{\tilde{1}\tilde{1}}\}$. The unitary transformations $\mathbb{L}$, $\mathbb{S}$ and $\mathbb{R}$ can be rewritten as
\begin{align}
&\mathbb{L}=\ket{\tilde{0}}\bra{\tilde{0}}\otimes L+\ket{\tilde{1}}\bra{\tilde{1}}\otimes L',\nonumber\\
&\mathbb{S}=S\otimes\ket{\tilde{0}}\bra{\tilde{0}}+\mathbb{I}\otimes\ket{\tilde{1}}\bra{\tilde{1}},\nonumber\\
&\mathbb{R}=\ket{\tilde{0}}\bra{\tilde{0}}\otimes R+\ket{\tilde{1}}\bra{\tilde{1}}\otimes R'.
\end{align}
Then the $4\times4$ unitary transformations $U(k\Delta t)$ can be implemented by these three controlled two-qubit transformations in \autoref{fig:circuit}.

\begin{figure}
  \includegraphics[width=0.45\textwidth]{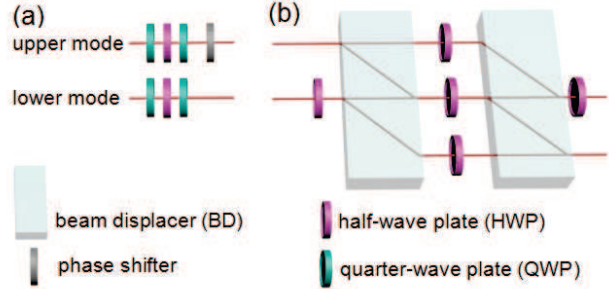}\\
  \caption{Conceptual experimental setup, with three controlled unitary transformations $\mathbb{L}$, $\mathbb{S}$ and $\mathbb{R}$. Red lines represent the optical modes (beams) of single photons. (a) Realization of $\mathbb{L}$ ($\mathbb{R}$) as transformation on two spatial modes and two polarizations modes of single photons. The spatial mode works as the control qubit and the $2\times2$ unitary transformation $L$ ($R$) and $L'$ ($R'$) applied on the polarizations of the photons in different modes can be realized by a set of wave plates (WPs). A phase shifter is used to keep the global phase unchanged during the transformation. (b) Realization of $\mathbb{S}$. The polarization is the control qubit. After the first HWP at $45^\circ$ and a BD, the horizontally polarized photons in both spatial modes are propagating in the same spatial mode (the middle one) and then the transformation $S$ is applied by using a HWP at $\theta/2$ in the middle mode. Meanwhile, the vertically polarized photons in upper or lower modes are not affected, and after the second BD they are still propagating in upper or low modes. The other HWPs are all set to $45^\circ$, which are used to flip the polarizations and to change the propagating modes of photons after they pass through the BD.}
  \label{fig:setup}
\end{figure}

\begin{figure}
  \includegraphics[width=0.5\textwidth]{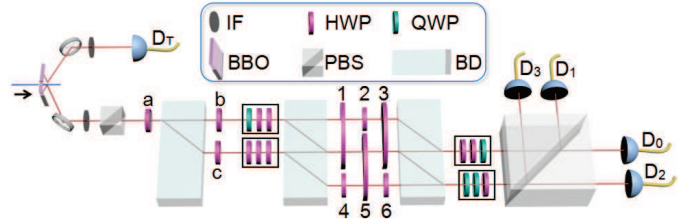}
  \caption{Practical experimental setup with consideration of both compensation of optical delay between different spatial modes and simplification.}
  \label{fig:s}
\end{figure}

\begin{figure}
  \includegraphics[width=0.45\textwidth]{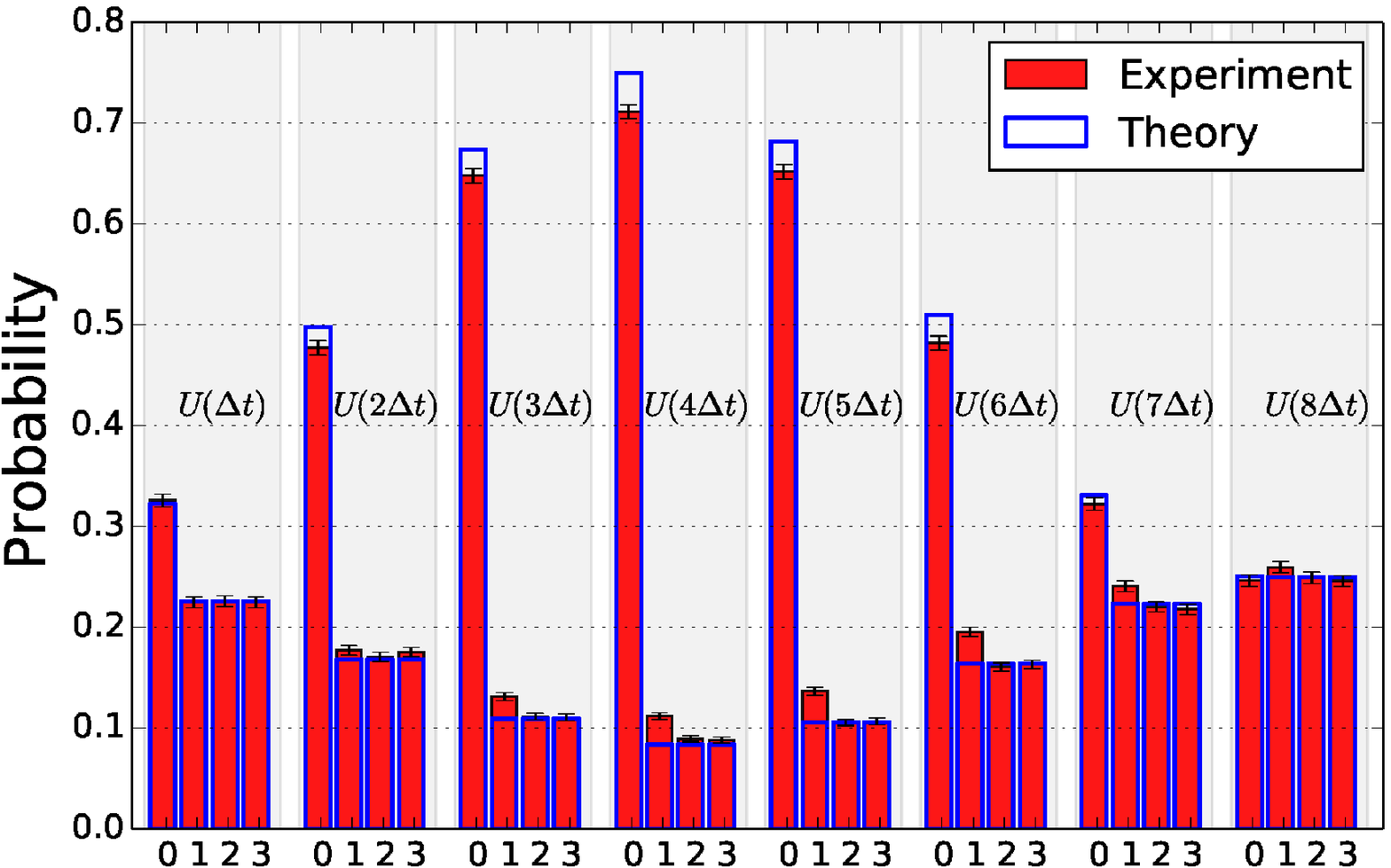}
  \caption{Photon probability distributions after eight unitary transformations. Red bars represent the experimental results. Blue borders represent the theoretical predictions. Errors are estimated via propagated Poissonian statistics. Small additional uncertainty may be present in the measurement of nodes 0 and 1 due to the photon representing the two states with different polarization but inhibiting the same spatial mode.}
\label{fig:result}
\end{figure}

The schematic of our experimental setup is depicted in \autoref{fig:setup}. The two qubits are encoded by spatial and polarization modes of single photons. The first qubit $\ket{\tilde{0}}$ ($\ket{\tilde{1}}$) represents upper (lower) spatial mode of photons, and the second qubit $\ket{\tilde{0}}$ ($\ket{\tilde{1}}$) represents the horizontal (vertical) polarization of photons.

Polarization-degenerated photon pairs are generated by type-\uppercase\expandafter{\romannumeral1} spontaneous parametric down-conversion (SPDC) in $0.5$mm-thick nonlinear-$\beta$-barium-borate
(BBO) crystal pumped by a $400.8$nm CW diode laser with $90$mW of power. The single photon is generated by triggering on the other photon. Interference filters (IFs) are used to restrict the photon bandwidth to $3$nm. The photons are in horizontal polarization after the first polarizing beam splitter (PBS). The initial state is
prepared by two steps. Firstly, after passing through a half-wave plate (HWP$_a$) at $22.5^\circ$ which rotates the polarization of single photons to equal superposition of horizontal and vertical polarizations, the photons are split into two parallel paths by a birefringent calcite beam displacer (BD) which transmits the
vertically polarized photons directly and displaces horizontally
polarized photons by $3$mm. Secondly, two HWPs (HWP$_b$ and HWP$_c$) at $-22.5^\circ$ and $22.5^\circ$ are inserted to the upper and
lower modes respectively to flip the polarizations. Thus
the state of the single photons is prepared in $\ket{\psi(0)}=\frac{1}{2}\sum_{j=0}^3\ket{j}$.

For the controlled two-qubit transformation $\mathbb{L}$ and $\mathbb{R}$, the spatial mode of photons serves as the control qubit and the polarization is the target qubit. In the upper and lower modes, the $2\times2$ unitary transformations $L$ ($R$) and $L'$ ($R'$) are applied on the polarization degrees of freedoms, which can be realized by a combination of QWPs and HWPs sequence inserted in the corresponding spatial mode.

For the $4\times4$ unitary transformation $\mathbb{S}$, the polarization of photons serves as the control qubit. A HWP at $45^\circ$ inserted in the lower input mode flips the polarizations of photons. After the first BD, the horizontally polarized photons in both of the upper and lower input modes are propagating in the same path (the middle one) and then the transformation $S$ is applied by using a HWP at $\theta/2$ in the middle path. Hence $S\otimes\ket{\tilde{0}}\bra{\tilde{0}}$ is applied on $\ket{\tilde{0}\tilde{0}}$ and $\ket{\tilde{1}\tilde{0}}$. The vertically polarized photons in upper or lower input modes are not affected and after the second BD they are still propagating in upper or lower output modes. That is $\mathbb{I}\otimes\ket{\tilde{1}}\bra{\tilde{1}}$ is applied on $\ket{\tilde{0}\tilde{1}}$ and $\ket{\tilde{1}\tilde{1}}$. Two HWPs at $45^\circ$ inserted in the other two paths (the propagating paths between two BDs) are used to flip the polarizations of photons in the paths and then the propagating modes change after the photons pass through the following BD. After the second BD, a HWP at $45^\circ$ is inserted in the upper output mode to compensate the effect of the first HWP in the lower input mode and flip the polarizations of photons back.

Our actual experimental setup is shown in \autoref{fig:s}, which takes into consideration of the compensation of optical delay between different spatial modes. The simplified set of wave plates (WPs) for the realization $\mathbb{L}$ and $\mathbb{R}$ of the eight $4\times 4$ unitary transformations $U(k\Delta t)$ are shown in \autoref{table:1}, including the setting angles of WPs. Two BDs and six HWPs (HWP$_1$-HWP$_6$) are used to realize $\mathbb{S}$ and compensate the optical delay. The setting angles of the HWP$_1$-HWP$_6$ are given in \autoref{table:2}.

In order to implement the proposed centrality algorithm experimentally, we discretise the CTQW time-evolution operator $U(t)=e^{-iLt}$ given by \autoref{eq:unitary} for the 4-vertex star graph, using 8 time steps of $\Delta t=9/40$ to ensure we sample the probability distribution adequately over one period (note that $T=\pi/\sqrt{3}\approx 8\Delta t$). 
After applied the unitary time-evolution operator $U(k\Delta t)$ with $k \in \{1, 2~...~8\}$, the quantum state is measured by a two-qubit projective measurement. A PBS is used to perform the projective measurement on the photons with the computational basis $\{\ket{\tilde{0}\tilde{0}},\ket{\tilde{0}\tilde{1}},\ket{\tilde{1}\tilde{0}},\ket{\tilde{1}\tilde{1}}\}$. The photons are detected by avalanche photon-diodes in coincidence with the trigger with a coincident window of $3$ns. The clicks of detectors D$_0$, D$_1$, D$_2$, and D$_3$ correspond to probabilities of the final state projected into the basis $\{\ket{\tilde{0}\tilde{0}},\ket{\tilde{0}\tilde{1}},\ket{\tilde{1}\tilde{0}},\ket{\tilde{1}\tilde{1}}\}$. We record the clicks for $5$s, and more than $18000$
coincidence counts are detected in an overall measurement time. The measured probability distributions are shown in \autoref{fig:result}, which is in excellent agreement with the theoretical predictions given by \autoref{eq:probability}. Here we use the norm-1 distance $d=\frac{1}{2}\sum_{x=0,1,2,3}|P^{exp}(x)-P^{th}(x)|$
 to evaluate the quality of experimental demonstration. For all eight $U$, we obtain $d_1=0.003$, $d_2=0.020$, $d_3=0.026$, $d_4=0.039$, $d_5=0.031$, $d_6=0.031$, $d_7=0.017$, and $d_8=0.009$. The distances are all smaller than $0.04$, which indicates successful experimental demonstrations of the $4\times 4$ unitary transformations.

\begin{table}[htbp]
  \centering
  \caption{The simplified sets of WPs with certain setting angles for realization of eight $4\times4$ unitary transformation. The subscript $k$ of $L$, $L'$, $R$ and $R'$ corresponds to the $k$th unitary transformation $U(k\Delta t)$. Q and H represent QWP and HWP respectively.}
  \label{table:1}
  \begin{tabular}{c|c|c||c|c|c}
    \hline
    $L/R$ & WPs & Angles ($^\circ$) & $L^\prime/R^\prime$ & WPs & Angles ($^\circ$)\\
    \hline
    $L_1$ & Q-H-H & 90,0,-3.3 & $L_1^\prime$ & H-H-H & 90,0,157.5\\
    $R_1$ & H-H-Q & 0,48.4,90 & $R_1^\prime$ & H-Q-Q & 0,22.5,22.5\\
    \hline
    $L_2$ & Q-H-H & 90,0,-6.7 & $L_2^\prime$ & H-H-H & 90,0,157.5\\
    $R_2$ & H-H-Q & 0,51.7,90 & $R_2^\prime$ & H-Q-Q & 0,22.5,22.5\\
    \hline
    $L_3$ & Q-H-H & 90,0,-10.5 & $L_3^\prime$ & H-H-H & 90,0,157.5\\
    $R_3$ & H-H-Q & 0,55.5,90 & $R_3^\prime$ & H-Q-Q & 0,22.5,22.5\\
    \hline
    $L_4$ & Q-H-Q & 90,14.9,0 & $L_4^\prime$ & H-H-H & 90,0,157.5\\
    $R_4$ & Q-H-H-Q & 0,0,30.1,0 & $R_4^\prime$ & Q-Q-Q-Q & 0,0,22.5,22.5\\
    \hline
    $L_5$ & Q-H-H & 90,0,-20.2 & $L_5^\prime$ & H-H-H & 90,0,157.5\\
    $R_5$ & H-H-Q & 0,65.2,90 & $R_5^\prime$ & H-Q-Q & 0,22.5,22.5\\
    \hline
    $L_6$ & Q-H & 90,0,26.8 & $L_6^\prime$ & H-H & 90,22.5\\
    $R_6$ & H-H-Q & 0,-18.2,90 & $R_6^\prime$ & H-Q-Q & 0,22.5,22.5\\
    \hline
    $L_7$ & Q-H-H & 90,0,-35.0 & $L_7^\prime$ & Q-Q-H & 0,0,112.5\\
    $R_7$ & H-Q & 100.0,90 & $R_7^\prime$ & H-H & 0,22.5\\
    \hline
    $L_8$ & Q-H-H-Q & 90,0,-44.4,0 & $L_8^\prime$ & Q-Q-Q-Q & -22.5,-22.5,90,90\\
    $R_8$ & Q-H-H-Q & 0,0,90.6,0 & $R_8^\prime$ & Q-Q-Q-Q & 0,0,22.5,22.5\\
    \hline
  \end{tabular}
\end{table}

\begin{table}[htbp]
  \centering
  \caption{The setting angles of HWP$_1$ and HWP$_2$ for realization of the eight $4\times4$ unitary transformations. The setting angles of HWP$_3$, HWP$_4$, and HWP$_5$ are set to be $-45^\circ$, and HWP$_6$ is set to be $45^\circ$ for all eight $U(k\Delta t)$.}
  \label{table:2}
  \begin{tabular}{c||c|c|c|c|c|c|c|c}
  \hline
     k & 1 & 2 & 3 & 4 & 5 & 6 & 7 & 8\\
     \hline\hline
     HWP$_1(^\circ)$ & 54.1 & 63.1 & 71.8 & 80.0 & 87.5 & 86.3 & 82.0 & 80.3 \\
     \hline
     HWP$_2(^\circ)$ & 144.1 & 153.1 & 161.8 & 170.0 & 177.5 & 176.3 & 172.0 & 170.3\\
     \hline
  \end{tabular}
\end{table}

\section{Conclusion}
\label{sec:conc}
In this study, we have proposed a CTQW-based quantum centrality algorithm, shown that it correlates well with classical measures, and verified its performance on general random graphs. The proposed quantum measure was then successfully implemented experimentally for the first time for a 4-vertex star graph. Notably, this algorithm requires an $N$-dimensional Hilbert space, compared to discrete-time quantum walk-based algorithms, which require $N^2$ dimensions for the same graph. Furthermore, this algorithm preserves the full quantum behavior of the walker, unlike the QSW, which mutes the quantum behaviour of the walker due to decoherence.

In our physical implementation of the proposed CTQW centrality algorithm,  the unitary operation of the walker on the graph is decomposed into unitary transformations on a two dimensional subspace, and realised by operating on the polarisation and spacial modes of single photons. This method can be used to decompose, in principle, any dimensional unitary operations into series of two dimensional unitary operations.  By making use of the coherent property of photons, the technology in our experiment is a competitive candidate for demonstrating arbitrary unitary operations, allowing it to be utilised for a wide array of quantum algorithms and quantum information processes.  This paper reports the first successful physical demonstration of a quantum centrality algorithm on a 4-vertex star graph.

\section{Acknowledgments}
This work has been supported by NSFC (No. 1174052 and 11674056), NSFJS (No. BK20160024), and Scientific Research Foundation of Graduate School of Southeast University (No. YBJJ1623). JAI would like to thank the Hackett foundation and The University of Western Australia for financial support.\\

\appendix

\section{Explicit expressions for $L$, $R$, $L'$ and $R'$}

For clarity, we provide here the explicit expressions for $L_k$, $R_k$, $L'_k$ and $R'_k$, where the subscript $k$ corresponds to the $k$th unitary transformation $U(k\Delta t)$.
\begin{align}
L_1=\left[\begin{array}{cc}
-0.9936i & -0.1132i\\
-0.1132 & 0.9936
\end{array}\right], L'_1=\left[\begin{array}{cc}
0.7071 & -0.7071\\
0.7071 & 0.7071\end{array}\right],\notag
\end{align}

\begin{align}
R_1=\left[\begin{array}{cc}
0.9936i & 0.1132\\
-0.1132i & 0.9936
\end{array}\right],
R'_1=\left[\begin{array}{cc}
0.7071 & 0.7071\\
-0.7071 & 0.7071\end{array}\right],\notag
\end{align}

\begin{align}
L_2=\left[\begin{array}{cc}
-0.9730i & -0.2307i\\
-0.2307 & 0.9730
\end{array}\right], L'_2=L'_1,\notag
\end{align}

\begin{align}
R_2=\left[\begin{array}{cc}
0.9730i & 0.2307\\
-0.2307i & 0.9730
\end{array}\right],
R'_2=R'_1,\notag
\end{align}

\begin{align}
L_3=\left[\begin{array}{cc}
-0.9341i & -0.3569i\\
-0.3569 & 0.9341
\end{array}\right], L'_3=L'_1,\notag
\end{align}

\begin{align}
R_3=\left[\begin{array}{cc}
0.9341i & 0.3569\\
-0.3569i & 0.9341
\end{array}\right],
R'_3=R'_1,\notag
\end{align}

\begin{align}
L_4=\left[\begin{array}{cc}
-0.8686i & 0.4955\\
-0.4955 & 0.8686i
\end{array}\right], L'_4=L'_1,\notag
\end{align}

\begin{align}
R_4=\left[\begin{array}{cc}
0.8686i & 0.4955\\
-0.4955 & -0.8686i
\end{array}\right],
R'_4=R'_1,\notag
\end{align}

\begin{align}
L_5=\left[\begin{array}{cc}
-0.7618i & -0.6478i\\
-0.6478 & 0.7618
\end{array}\right], L'_5=L'_1,\notag
\end{align}

\begin{align}
R_5=\left[\begin{array}{cc}
0.7618i & 0.6478\\
-0.6478i & 0.7618
\end{array}\right],
R'_5=R'_1,\notag
\end{align}

\begin{align}
L_6=\left[\begin{array}{cc}
0.5926i & 0.8055i\\
0.8055 & -0.5926
\end{array}\right], L'_6=L'_1,\notag
\end{align}

\begin{align}
R_6=\left[\begin{array}{cc}
0.5926i & 0.8055\\
0.8055i & -0.5926
\end{array}\right],
R'_6=\left[\begin{array}{cc}
-0.7071 & -0.7071\\
-0.7071 & 0.7071\end{array}\right],\notag
\end{align}

\begin{align}
L_7=\left[\begin{array}{cc}
0.3415i & -0.9399i\\
0.9399 & 0.3415
\end{array}\right], L'_7=L'_1,\notag
\end{align}

\begin{align}
R_7=\left[\begin{array}{cc}
0.3415i & 0.9399\\
-0.9399i & 0.3415
\end{array}\right],
R'_7=R'_6,\notag
\end{align}

\begin{align}
L_8=\left[\begin{array}{cc}
0.0207i & -0.9998\\
0.9998 & -0.0207i
\end{array}\right], L'_8=L'_1,\notag
\end{align}

\begin{align}
R_8=\left[\begin{array}{cc}
0.0207i & 0.9998\\
0.9998 & 0.0207i
\end{array}\right],
R'_8=R'_6.\notag
\end{align}

\bibliography{unitary4x4}

\end{document}